\documentclass[prb,twocolumn,showpacs,amsmath,amssymb,superscriptaddress]{revtex4}
\usepackage[pdftex]{graphicx}
\usepackage{epsfig}

\begin{document}
\title{Elucidation of the helical spin structure of FeAs}
\author{T. Frawley}
\affiliation{Durham University, Department of Physics, South Road,
  Durham, DH1 3LE, UK}
\author{R. Schoonmaker}
\affiliation{Durham University, Department of Physics, South Road,
  Durham, DH1 3LE, UK}
\author{S.H. Lee}
\author{C.-H. Du}
\affiliation{Physics Department, Tamkang University, Tamsui 251, Taiwan}
\author{P. Steadman}
\affiliation{Diamond Light Source, Harwell Science and Innovation
  Campus, Didcot, Oxon, OX11 ODE}
\author{J. Strempfer}
\affiliation{DESY Photon Science, Notkestrasse 85, 22607, Hamburg, Germany}
\author{Kh. A. Ziq}
\affiliation{King Fahd University of Petroleum and Minerals, Department of Physics, Dhahran 31261, Saudi Arabia}
\author{S.J. Clark}
\affiliation{Durham University, Department of Physics, South Road,
  Durham, DH1 3LE, UK}
\author{T. Lancaster}
\affiliation{Durham University, Department of Physics, South Road,
  Durham, DH1 3LE, UK}
\author{P.D. Hatton}
\affiliation{Durham University, Department of Physics, South Road,
  Durham, DH1 3LE, UK}

\begin{abstract}
We present the results of resonant x-ray scattering measurements and
electronic structure calculations on the monoarsenide FeAs. We
elucidate details of the magnetic structure, showing the ratio of
ellipticity of the spin helix is larger than previously thought, at
2.58(3), and reveal both a right-handed chirality and an out of plane
component of the
magnetic moments in the spin
helix. We find that electronic structure calculations and analysis of the
spin-orbit interaction are able to qualitatively account for this canting. 
\end{abstract}
\maketitle

\section{Introduction}

\begin{figure}
\begin{center}
\includegraphics[width = \columnwidth]{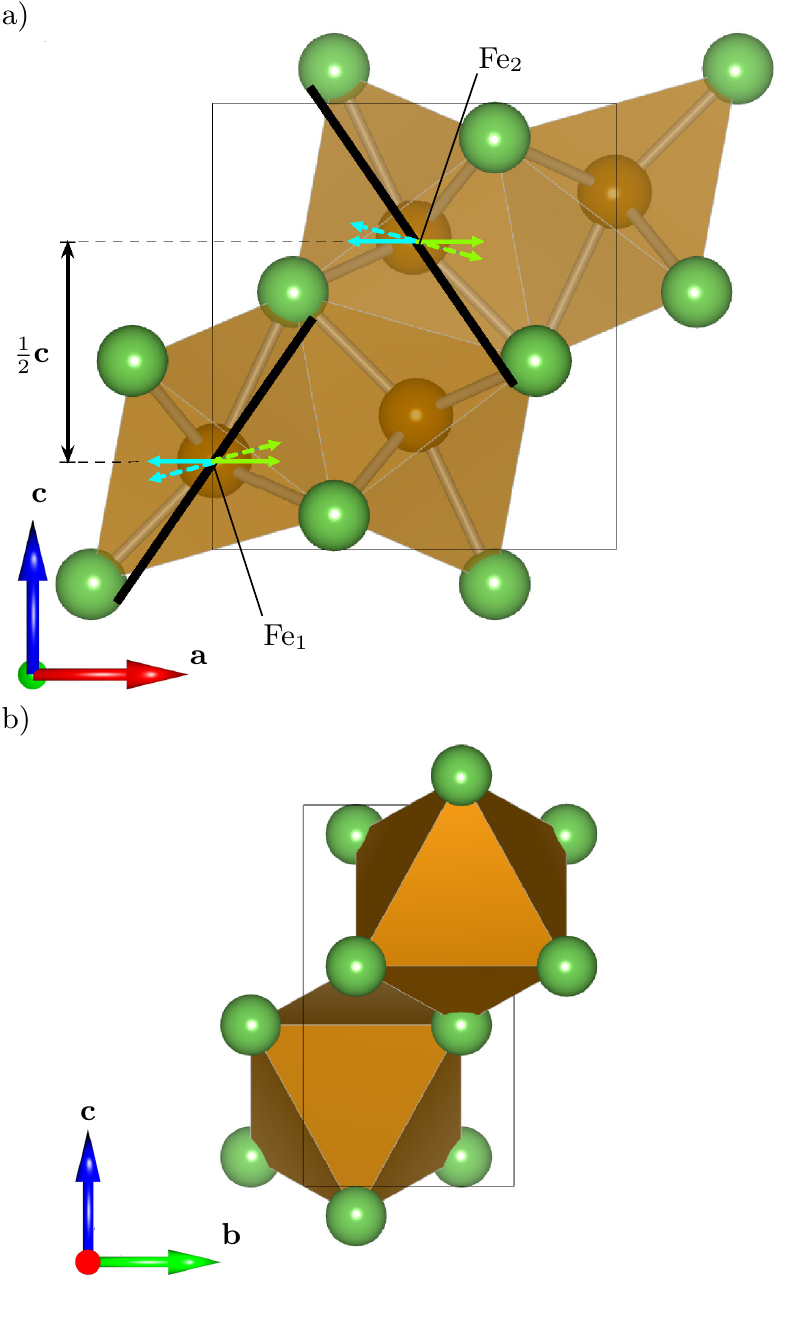}
\caption[FeAs structure.]{Crystal structure of FeAs in 
(a) the $a$-$c$ plane; and (b) the $b$-$c$ plane. (As atoms denoted by green circles.)}
\label{fig:feas:cryst}
\end{center}
\end{figure}

Unlike the
cuprates, where the magnetic state owes its existence to Mott insulator
physics, in iron-based superconductors magnetism results
from an instability of the delocalised Fe $d$-band electrons 
which gives rise to a spin-density wave
\cite{basov,ishida,norman,nandi,mazin}. The pnictide parent compounds
 display 
 metallic, antiferromagnetic spin-density wave ground states
where the spins are periodically modulated in 
space but where the outermost electrons can be delocalised, 
 typical of the collective effect that emerges from
 an instability of the paramagnetic Fermi surface.   A key question remains how the parent
magnetic state in pnictides evolves across the phase diagram and how the properties
of the doped magnet can
compete, promote or coexist with superconductivity. 
The simplest of all iron arsenide systems, the monoarsenide FeAs, may
provide some insights into these questions since, it has been argued \cite{rodriguez}, its itinerant
magnetism is related to the magnetic ground states of iron-based 
superconductors.

\begin{figure}
\begin{center}
\includegraphics[width = \columnwidth]{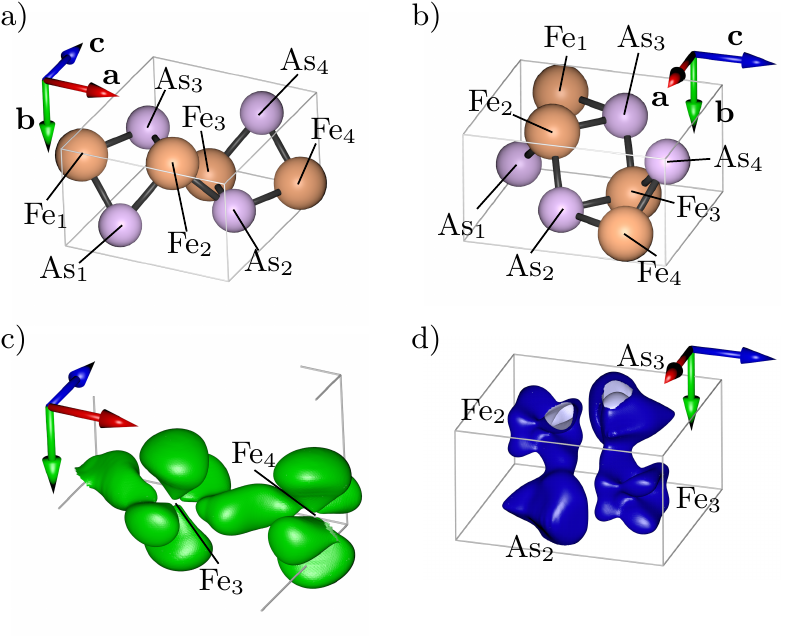}
\caption{(a) and (b) Crystal structure of FeAs showing the different
  Fe atoms. Selected regions of (c) electron density decrease and (d)
  density increase on change from zero-spin to AFM1 calculated
  states. \label{fig:robert1}}

\end{center}
\end{figure}

FeAs crystallises in the B31 (MnP-type) structure (space group $Pnma$) 
\cite{FeAs:Selte:1969} 
which consists of distorted FeAs$_{6}$ octahedra
which are face sharing along the $a$-axis and edge sharing along the
$b$- and $c$-axes (Fig.~\ref{fig:feas:cryst}). 
It therefore has similar Fe-Fe linkages to the layered Fe-based
superconductors (such as LaFeAsO, BaFe$_{2}$As$_{2}$ and NaFeAs), but
is distinguished from them by being surrounded
by six (octahedral) rather than four (tetrahedral) arsenic atoms. 
The iron atoms sit at the \textit{4c} Wyckoff site,  giving rise to
four positions in the unit cell: Fe$_{1}$ at\cite{rodriguez}  (\textit{x}, $\frac{1}{4}$,
\textit{z}), Fe$_{2}$ at  (\textit{$\bar{x}$}+$\frac{1}{2}$,
$\frac{3}{4}$,\textit{z}+$\frac{1}{2}$), Fe$_{3}$ at (\textit{$\bar{x}$},
$\frac{3}{4}$, \textit{$\bar{z}$})  and Fe$_{4}$ at
(\textit{x}+$\frac{1}{2}$, $\frac{1}{4}$,
\textit{$\bar{z}$}+$\frac{1}{2})$, where $x=0.004$ and $z=0.199$
as shown in Fig.~\ref{fig:robert1}.

Initial neutron powder diffraction  measurements \cite{FeAs:Selte:1972}
showed that FeAs undergoes
a transition to a long-range 
antiferromagnetically ordered state at $T_{\mathrm{N}}=77$~K. It was
suggested that the system
adopts
a helical magnetic structure with a wavevector 
$q=[0,0,0.375]$ and an ordered magnetic moment of 0.5~$\mu_{\mathrm{B}}$.  
However, a more recent magnetic susceptibility and transport study 
\cite{segawa}, gave results that
indicated the presence of significant magnetocrystalline anisotropy,
raising 
doubts about the occurrence of such a simple spin helix structure. 
Specifically, single crystal susceptibility shows a kink at 70~K in the
$a$- and $b$-directions but not in the
$c$-direction.  (The lack of features in the $c$-axis direction
suggesting the magnetic moment is fixed in the $a$-$b$
plane.)
The key observation is that the susceptibility along the $b$-axis is
lower than that along $a$, and only the $b$-axis displays a
magnetic field splitting, suggesting the presence of anisotropy.
Following this, a polarised single crystal neutron
diffraction study \cite{rodriguez} suggested a slightly elliptical helical structure 
comprising a non-collinear spin-density wave arising from a
combination of itinerant and localised behaviour, with the spin amplitude along the $b$-axis direction being
1.5(5)\% larger than that along the $a$-direction. 

In terms of its electronic properties, FeAs lies
between two well-understood regimes: the delocalized magnetic metal and
the localized magnetic insulator. 
Resistivity measurements,\cite{segawa} confirm the
itinerant behaviour of FeAs: resistivity decreases below
150~K with a kink observed at 70~K.  
Electronic structure calculations on FeAs have not fully
elucidated the mechanism that generates the magnetic
structure. Non-polarized, collinear and non-collinear spin
calculations have been carried out, but
find the lowest energy state to be antiferromagnetic, in which
nearest-neighbour iron spins antialign with a resultant $P2_1/m$
symmetry \cite{mazin,griffin2012,griffin2014}. 
The
study by Parker and Mazin \cite{mazin} calculated the static Lindhard function in
FeAs and the nesting of the Fermi surface in the AFM phase,
and concluded that some form of nesting did not drive the magnetic
order. Griffin and Spaldin \cite{griffin2014} compared the use of
different DFT functionals in FeAs, and found that GGA/hybrid
GGA gives values for the structure in closest agreement with experiment, but that a
negative Hubbard-$U$ calculation  would be most
likely to reproduce the spin spiral as it would increase competition
between AFM and FM interactions and increase the energy of the AFM
state. However, there is no other physical justification for a negative
Hubbard-$U$ parameter, implying a larger failing in these
functionals. Griffin and Spaldin also
performed non-collinear spin calculations imposing a variety of spirals on
the system, but found that the AFM state was lower in energy than all
of them.

In this paper we present a 
refinement of the magnetic structure of iron arsenide using x-ray
resonant elastic scattering (XRES) and calculations of the electronic
structure
using density functional theory (DFT). From analysis of a magnetic
satellite reflection, our XRES results 
strongly suggest that the magnetic spiral is considerably more
elliptical than was previously believed,  has a
right-handed chirality and has an ordered spin component in
the propagation direction of the helix. 
DFT calculations show that spin-orbit interactions and the local iron environment
provide an explanation for this new
ordering component. 
We conclude that the spin ordering is linked to localized
orbital restructuring and changes in electronic density, and therefore, as might
be expected, is not well described by simple Stoner or Ising-type models.

\section{Experimental and computational methods}

Our sample of FeAs was grown by an iodine vapour transport
method\cite{ziq}. The growth method resulted in single crystals of
typical dimensions $\approx 100~\mu$m. 
Several samples were characterized using a four-circle
diffractometer, and a single crystal selected, with a natural
\textit{c}-axis facet and a sharp $[0,0,2]$ reflection with a rocking
width of just $0.0025^\circ$.   

XRES measurements were carried out at both the
soft Fe $L_\textrm{II/III}$ and the hard Fe \textit{K} absorption
edges.  For the Fe \textit{L} edges experiments the beamlines ID08 (at
ESRF) and I10 (at Diamond) were used.  For the \textit{K} edge experiments
the beamline P09 (at Petra III) was used. All
three beamlines are situated on an undulator insertion device.  
For the ID08 and I10 experiments the
sample was mounted with the $b$-axis in the scatter plane.  For
the P09 experiment the $[-1,0,0]$ reciprocal direction was used as the
azimuthal reference vector.  
  
DFT calculations were run with the CASTEP electronic structure code using the
PBE exchange-correlation
functional\cite{DFTgen}\cite{CASTEP}. Energy
differences between spin configurations were converged to 1 part in
10,000, and to generate the Fermi-surface a Monkhorst-Pack $k$-point
grid of $23 \times 27 \times 19$ was used. To account for core
state contributions on atoms an ultrasoft core-corrected iron
pseudopotential with 8 valence electrons and an arsenic
pseudopotential with 15 valence electrons were used. A non-magnetic
configuration and
a range of collinear ordered spin-structures were considered.

\section{X-ray scattering results}

\subsection{Soft x-ray scattering}

\begin{figure}[!t]
\begin{center}
\includegraphics[width = \columnwidth]{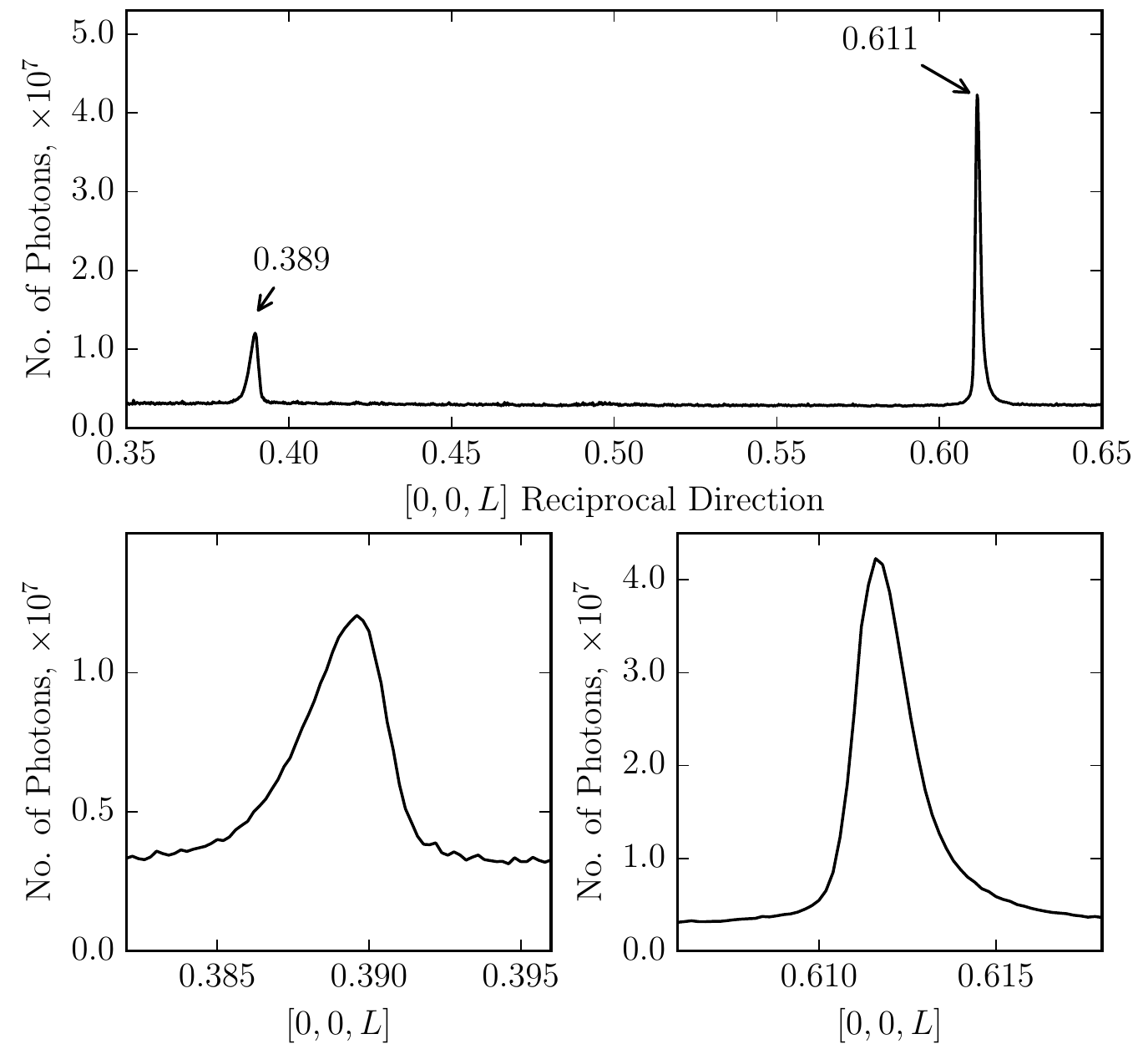}
\caption[Reciprocal space scan along ${[00L]}$ at Fe
L${_\textrm{III}}$]
{Scan along the \textit{L} reciprocal lattice direction, at the Fe
  $L_\mathrm{III}$ edge.  
Due to the large wavelength of the Fe $L_\mathrm{III}$ edge the Ewald sphere is limited to 0.68$c^{*}$.}
\label{fig:FeAs:L3lscan}
\end{center}
\end{figure}

\begin{figure}[!t]
\begin{center}
\includegraphics[width = \columnwidth]{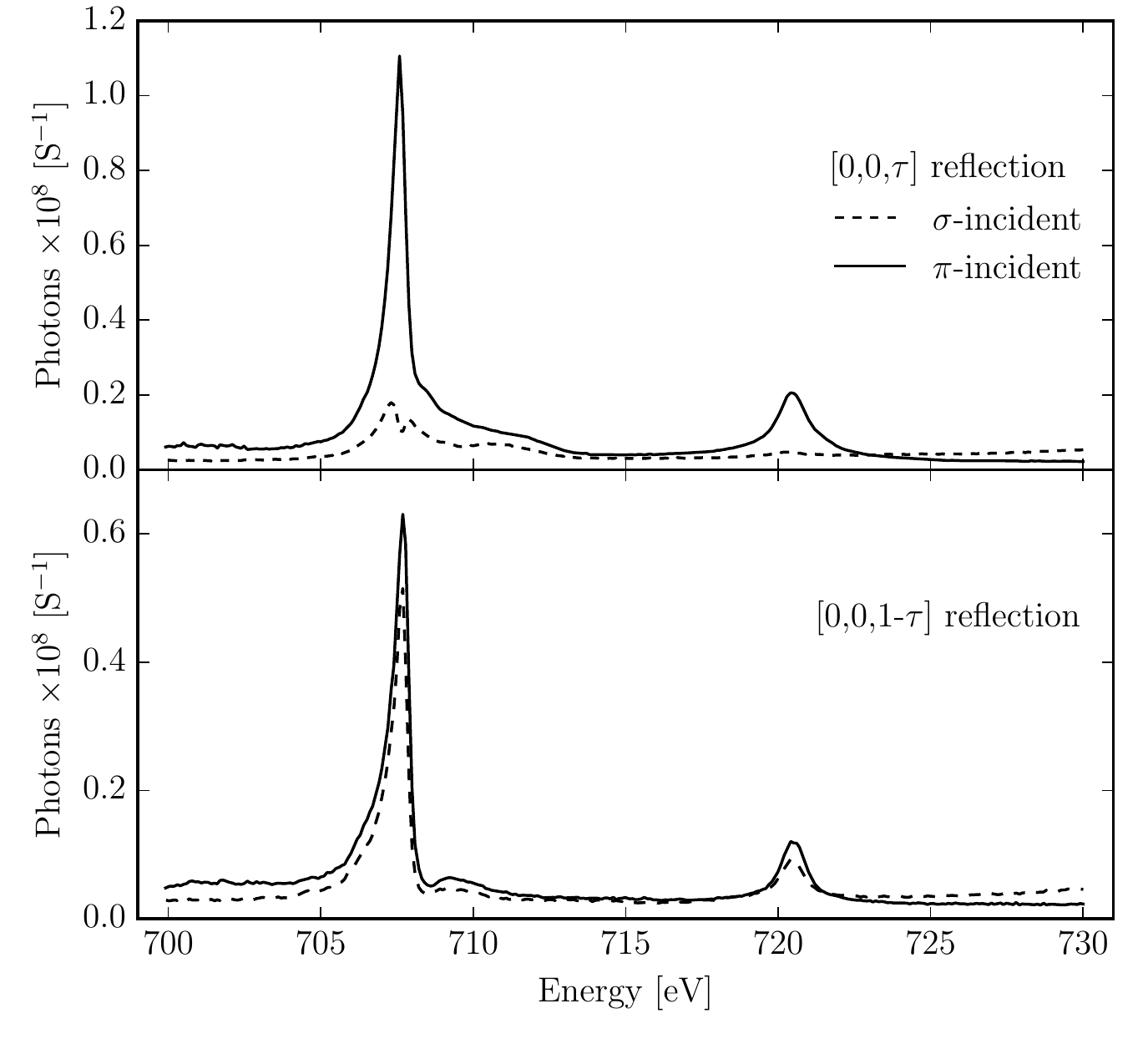}
\caption[Energy resonance of magnetic Bragg peaks at Fe
L${_\textrm{III}}$ (linear polarisation)]{Energy resonance of the
  magnetic satellite peaks, $[0,0,\tau]$ (top) and $[0,0,1-\tau]$
  (bottom), with $\sigma$- and $\pi$-polarised incident light.  No
  post-scatter polarisation analysis was used. }
\label{fig:FeAs:LINenergySOFT}
\end{center}
\end{figure}

At the Fe $L_{\textrm{III}}$ energy ($\approx 707$~eV) the radius of
the Ewald sphere limits access along the $l$ reciprocal
direction to $l=0.68$. Within this limit two resonant reflections were
found at positions $l=0.389$ and $l =0.611$.  A scan along the [00$L$] direction is shown in 
Fig.~\ref{fig:FeAs:L3lscan}.  The observed peaks are asymmetric (most
likely due to the energy profile of the undulator), with
the peak at $l=0.611$ having the reverse asymmetry to the $l=0.389$
reflection.  
This suggests that
[0,0,0.611] is a
satellite of the forbidden $[0,0,1]$
Bragg peak.  The two reflections can be indexed as
$[0,0,0]+[0,0,\tau]$ (denoted $\tau$ hereafter), and
$[0,0,1]-[0,0,\tau]$ (denoted $1-\tau$ hereafter), where $\tau=0.389$.
Energy resonances of the two reflections, without post-scatter
polarisation analysis (Fig.~\ref{fig:FeAs:LINenergySOFT}),
  were performed by
decreasing the energy of the incident x-ray whilst maintaining the
diffraction condition for the magnetic peak.  The resonances were
measured with both $\sigma$- and $\pi$-polarised incident light.
 Assuming a dipole-dipole transition ($E1E1$) is responsible for the
resonant feature,  the transition is from the Fe 2$p$ orbital
to the Fe 3$d$ band.  Exciting into this Fe 3$d$ band leads to
the sensitivity of the technique to the local magnetism, as it is the $3d$
orbitals that are the magnetically active spin-polarised band in
iron. 
We also note that the temperature behaviour of the [0,0,$\tau$]
magnetic Bragg peak shows critical 
behaviour consistant with that observed previously\cite{rodriguez}.

The $\tau$ reflection shows a marked difference between
the two polarization channels, which is sufficient to rule out either charge scattering or
simple collinear magnetic spin structures along the $a$-, $b$-, or
$c$-directions, assuming a $E1E1$ origin to the
scattering.
The $1-\tau$
reflection shows very different behaviour, giving equal intensity
with incident $\sigma$- and $\pi$-polarised light (Fig.~\ref{fig:FeAs:LINenergySOFT}).  This indicates a
different origin for the two peaks.  The $\tau$ and
$1-\tau$ reflections occur at $\theta$ angles of
$34.4^{\circ}$ and $63.7^{\circ}$, respectively.  These angles are not
close to $45^\circ$ or $90^\circ$, which might cause a suppression of
scattering due to the $\theta$ dependences of the scattering
amplitude.
For the $\tau$ reflection
the intensity in the circular-positive channel is roughly twice that
of the circular-negative channel,  indicative of a
noncollinear spin structure.  The $1-\tau$ reflection has
equal intensity in the circular-positive and circular-negative
channels. This again shows very different behaviour to the
$\tau$ peak.

\begin{figure}[!t]
\begin{center}
\includegraphics[width = \columnwidth]{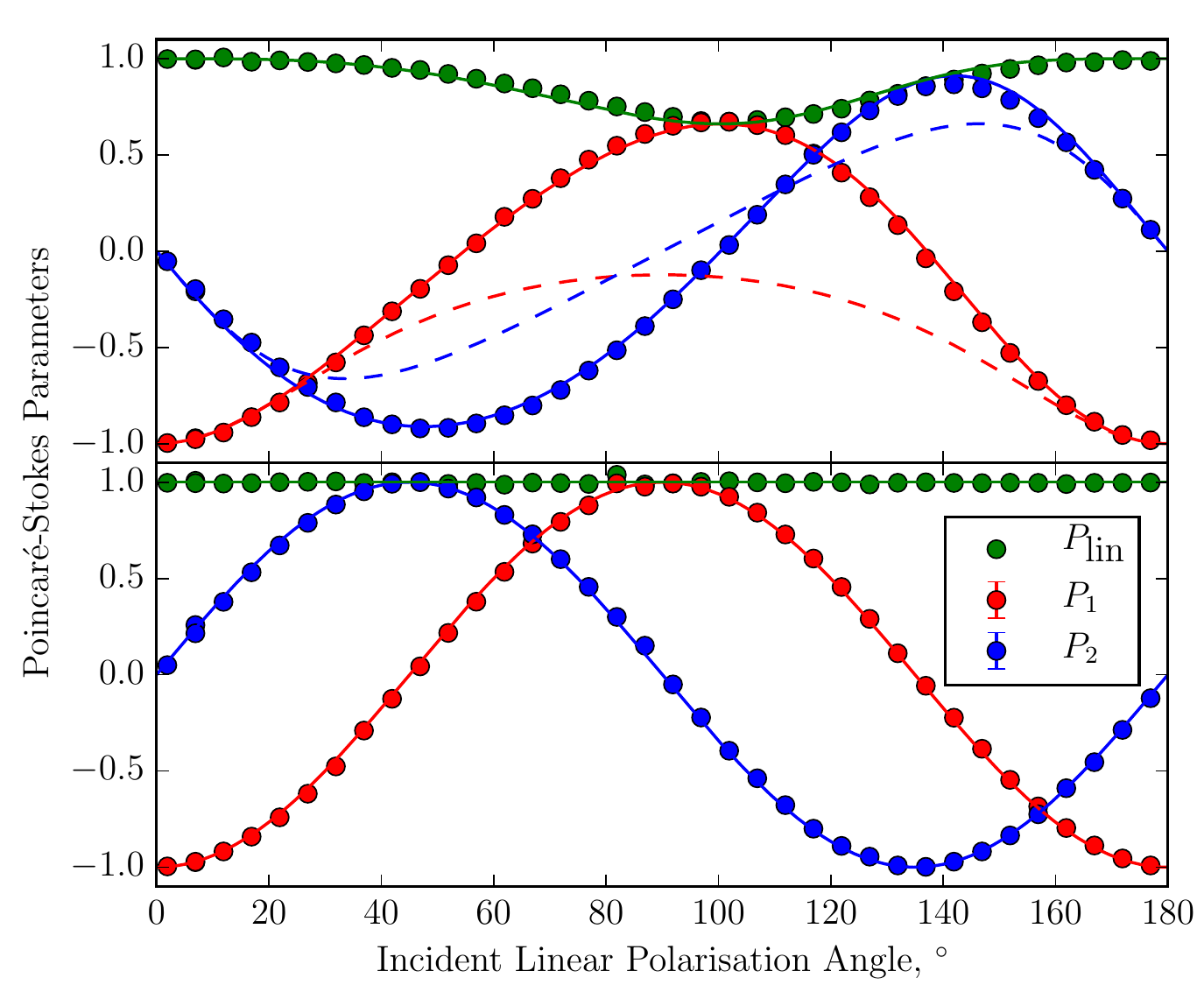}
\caption[FLPA measurement of ${[0,0,\tau]}$ and ${[0,0,1-\tau]}$]{Full
  linear polarisation analysis (FLPA).
Top: FLPA measured on the
  $[0,0,\tau]$ reflection. Bottom: FLPA measured on the
  $[0,0,1-\tau]$ reflection. The solid lines are the elliptical helix model based on the
  derived structure factor. The dashed lines are the predictions for a
  circular helix discussed in Section 4c.}
\label{fig:FeAs:Softpolanalysis}
\end{center}
\end{figure}

A full linear
polarisation analysis (FLPA) was also carried out on both peaks
using the ultra-high vacuum diffractometer,
RASOR at I10,
 and the results
are shown in Fig.~\ref{fig:FeAs:Softpolanalysis}.  In this measurement
the incident linear light is rotated through a full 180$^\circ$, and
at each incident polarisation angle, the polarisation state of the
scattered beam is measured. Fig.~\ref{fig:FeAs:Softpolanalysis} shows
the incident polarisation angle against the outgoing polarisation
using Poincar\'e-Stokes parameters, $P_1$ and $P_2$.
The results show a different polarisation analysis for the
$\tau$ and $1-\tau$ reflections, confirming
they have  different origins.

\subsection{Hard x-ray scattering}

XRES measurents at the hard x-ray energy
of the Fe \textit{K} absorption edge allow
for a wider field of access to reciprocal space than measurements at the soft energies,
 but at a cost of sensitivity to the magnetism.  A typical
$E1E1$ transition at the Fe \textit{K} absorption edge, excites a Fe
$1s$ electron into the empty Fe $4p$ band.  The sensitivity to
magnetism arises from any overlap, or hybridisation, between the Fe
$3d$ and Fe $4p$ bands.

\begin{figure}[!t]
\begin{center}
\includegraphics[width = \columnwidth]{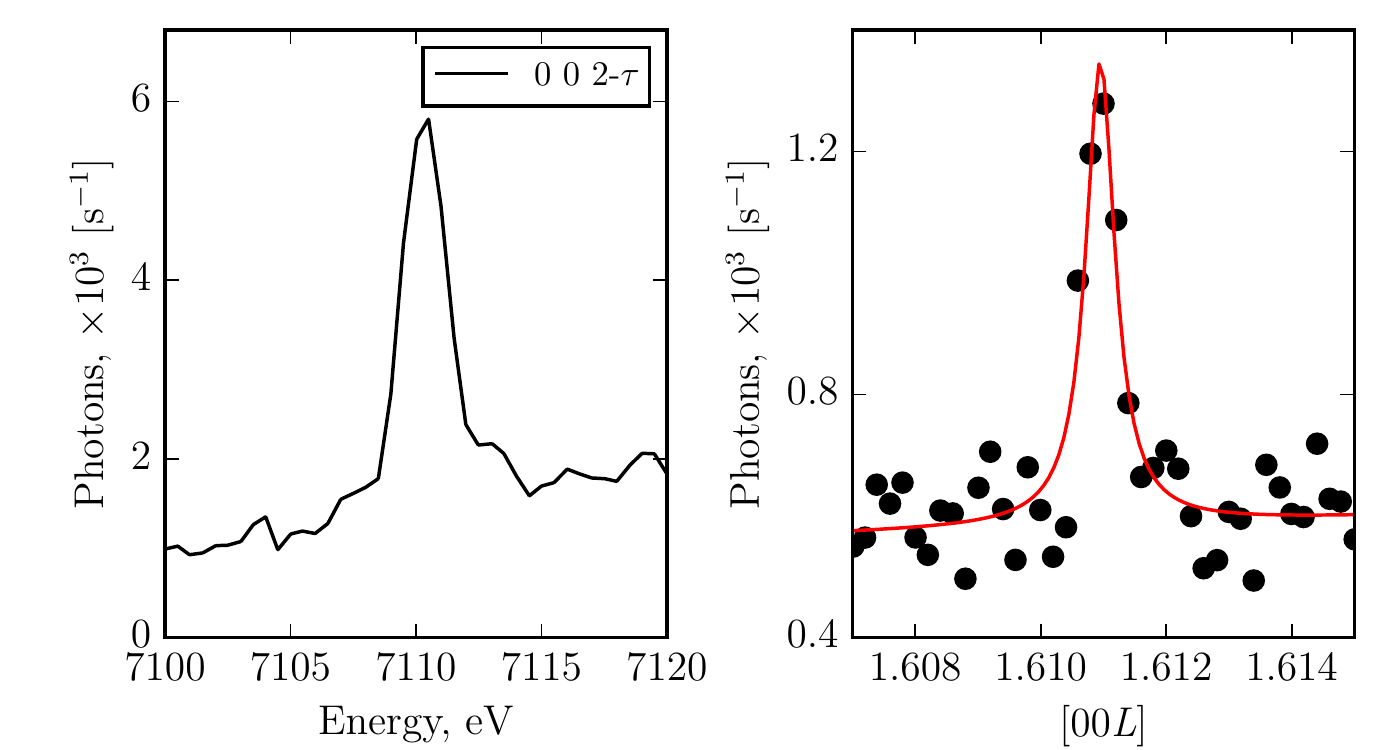}
\caption[Energy resonance of ${[0,0,2-\tau]}$ at Fe K edge]{Left:
  energy scan of resonance of the $[0,0,2-\tau]$ magnetic reflection.  
Right: $[00L]$ Reciprocal space scan of $[0,0,2-\tau]$. Scans were performed in the $\sigma-\pi$ channel.}
\label{fig:FeAs:resTAU}
\end{center}
\end{figure}

A survey of resonant reflections was carried out and satellite
reflections were found at $[0,0,2-\tau]$, $[0,0,2+\tau]$,
$[0,0,4-\tau]$, $[0,0,4+\tau]$, as well as at $[0,0,2\tau]$, and
$[0,0,3\tau]$.  An off-axis reflection was also observed at the
$[1,0,3-\tau]$ position.  No reflections were found at positions away
from the odd forbidden Bragg peaks ($[0,0,1\pm\tau]$ and
$[0,0,3\pm\tau]$).  
Figure~\ref{fig:FeAs:resTAU}
shows the resonances
and reciprocal space scans of the $[0,0,2-\tau]$ peak.
All types of satellite reflection show a sharp
resonant feature at 7110~eV.  The reciprocal space scans show the
$[0,0,2-\tau]$ peak to be the sharpest with a width of
$0.0006(1)$~r.l.u., while the $2\tau$ and $3\tau$ reflections
are wider with widths of $0.0019(1)$ and $0.0014(1)$
r.l.u. respectively. The $\tau$ and $2\tau$ reflections
were found only in the $\sigma-\pi$ channel and not in the
$\sigma-\sigma$ channel, while the $3\tau$ reflection was found in
both channels but was stronger in the $\sigma-\sigma$ channel. An
$E1E1$-type transition can produce $\tau$ and $2\tau$
reflections, but a quadrupole-quadrupole type transition (E2E2, involving an excitation from the $1s$ orbital into the
magnetically active $3d$ spin-polarised band) is
required to explain the presence of a $3\tau$ reflection. 
  
\begin{figure}[!t]
\begin{center}
\includegraphics[width = \columnwidth]{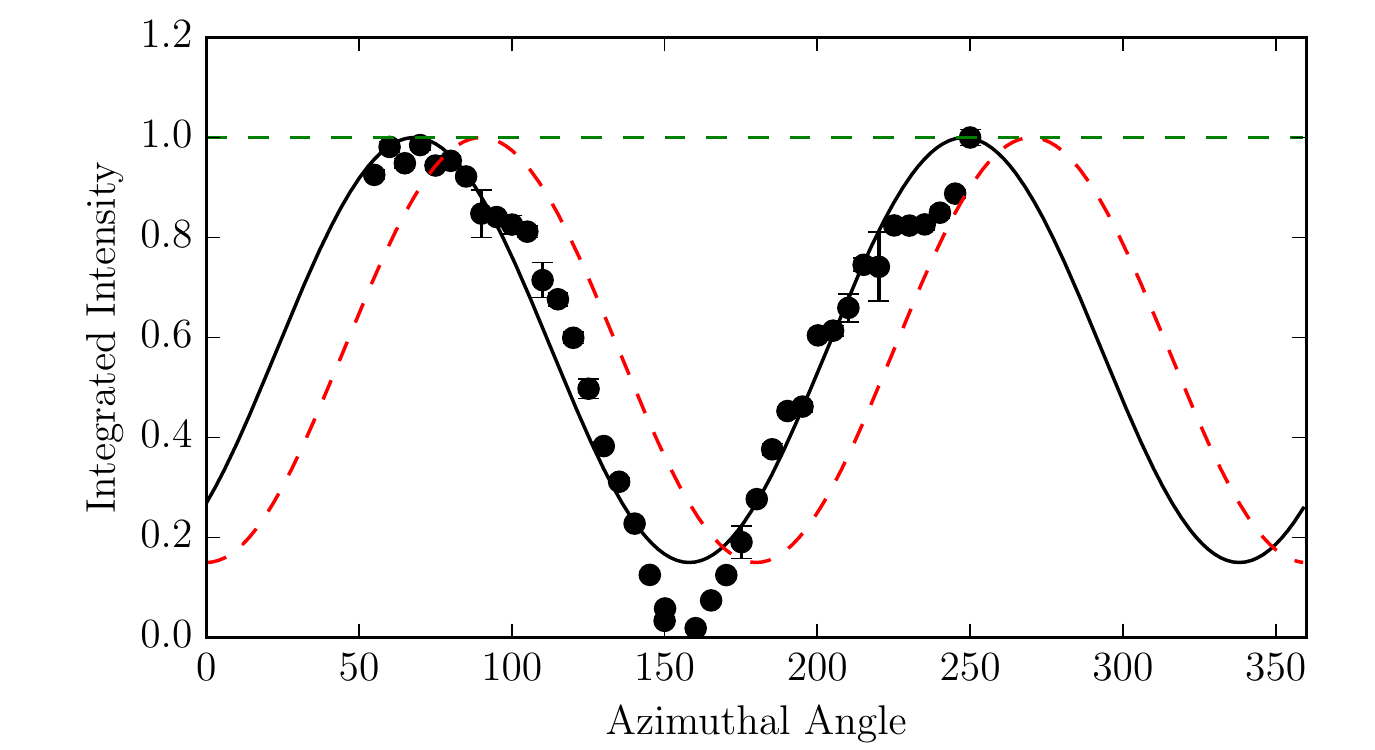}
\caption[Azimuthal measurement of the ${[0,0,2-\tau]}$ magnetic Bragg
peak.]{Azimuthal measurement of the $[0,0,2-\tau]$ magnetic Bragg
  peak.  The dashed and solid lines show predictions made using the
  structure factor.  The
  green dashed line shows the expected azimuth for circular helical
  magnetic structure.  The red dashed line shows the expected azimuth
  for a elliptical helical magnetic structure ($m_a:m_b = 1:2.58$).
  The  solid black line shows the prediction for the elliptical helix
  rotated by $-22^\circ$. }
\label{fig:FeAs:azimuthP09}
\end{center}
\end{figure}

An azimuthal measurement was performed on the $[0,0,2-\tau]$
reflection (Fig.~\ref{fig:FeAs:azimuthP09}) which involves a rotation of the sample around the scattering vector,
maintaining the diffraction condition. 
The zero point on the azimuthal axis is
defined as when the $[-1,0,0]$ reciprocal vector is in the scattering
plane away from the incident beam.  
Qualitatively, these azimuth data rules out a simple
non-elliptical helix. For the $[0,0,2-\tau]$ reflection, the
scattering vector is parallel to the magnetic propagation direction,
an azimuthal measurement rotates around the magnetic propagation
vector.  If the magnetic helix was circular, then there would be no
change in moment direction upon an azimuthal rotation, and constant
intensity would be expected.

The observation of not just the [0,0, $\tau$] but also the [0,0, 2$\tau$]  and
[0,0,3$\tau$] reflections is very reminiscent of x-ray scattering studies
of chromium metal. Chromium is the canonical example of an itinerant,
incommensurate antiferromagnet below its N\'{e}el temperature $T_{\mathrm{N}} = 311$~K
\cite{[14]}. This transverse spin density wave (SDW) is well understood and
arises from a nesting effect between the 3$d$ electron pocket centered
at the $\Gamma$ point and the hole pocket centered at the edge of the
Brillouin zone at the $H$ point. The pairing is between momentum states
separated by wavevector $Q$ and a spiral density wave is formed. The ground state
is then formed from two spiral waves of opposite helicity resulting in
a long period, linearly polarised SDW. Associated with the SDW
ordering, there is a distortion of the lattice with twice the
wavevector of the SDW, causing a charge density wave (CDW). This
results in differing satellites surrounding the Bragg peaks with odd
multiples of $Q$ being magnetic satellites resulting from the SDW and
even multiples of $Q$ caused by the CDW. Thus $\pm Q$ and $\pm 3Q$ magnetic
satellites were first observed by neutron diffraction \cite{[15]} and 
$\pm 2Q$ and
$\pm 4$ charge satellites observed by x-ray scattering \cite{[16]}. The
observation of the fourth harmonic suggests that the CDW is not
perfectly sinusoidal. It is not yet clear as to the exact mechanism
for producing a density wave in the charge distribution, with both a
magnetostriction effect (coupling the elasticity to the magnetism)
\cite{[17]}, or a purely electronic effect based on nesting between
electronic bands \cite{[18]}, being claimed. This is similar, but slightly
different to our observations, suggesting that in FeAs the $\tau$ and
$2 \tau$
satellites are magnetic in character but the $3 \tau$ satellite has both
magnetic and charge characteristics. Further studies of the
temperature dependence of these satellites may help unravel their
origin.

\section{Magnetic structure determination}
\subsection{The nature of the ellipse}
To determine the magnetic structure of the material, 
we consider an elliptical magnetic helix pointing along the $c$-axis 
with spin components $S_{a}$ and $S_{b}$
in the $a$- and $b$-directions respectively and a chirality $\chi(=\pm
)$. The magnetic moment on the $n$th atom in the $j$th unit cell for a
spin helix with propagation vector $\boldsymbol{\tau}$ is
given by
\begin{eqnarray}
\boldsymbol{m}_{n,j}(\boldsymbol{r}_{n,j}) &=& 
S_{a}\cos(\boldsymbol{\tau}\cdot \boldsymbol{r}_{n,j}-\psi_{n})\hat{\boldsymbol{a}}\\
&+&
S_{b}\cos(\boldsymbol{\tau}\cdot
\boldsymbol{r}_{n,j}-\psi_{n}+\chi\pi/2)\hat{\boldsymbol{b}}, 
\end{eqnarray}
where $\psi_{n}$ is the
phase shift caused by the orbit of Fe$_{n}$.
For such a helix propagating along the $[0,0,L]$ direction, satellite
peaks only appear around even Bragg reflections with structure factor
\begin{eqnarray}
f_{00l} &\propto &
  \left(\boldsymbol{\varepsilon}^\prime\times\boldsymbol{\varepsilon}\right)\cdot\mathbf{M}\left(e^{i2\pi
      z_{1} l}
e^{-i\psi_1}+ e^{-i2\pi  z_{1}
  l}e^{-i\psi_3}\right)
\nonumber
\\
 &+&
\left(\boldsymbol{\varepsilon}^\prime\times\boldsymbol{\varepsilon}\right)\cdot\mathbf{M}^*\left(e^{i2\pi
       z_{1} l}e^{i\psi_1}+ e^{-i2\pi  z_{1} l}e^{i\psi_3}
   \right),
\label{eqn:spinstructlSeven}
\end{eqnarray}
where $\boldsymbol{\varepsilon}^\prime\times\boldsymbol{\varepsilon}$ is
a polarization factor,
$\boldsymbol{M}=S_{a}\hat{\boldsymbol{a}}+\mathrm{i}\chi
S_{b}\hat{\boldsymbol{b}}$,
$z_{1}$ is the fractional coordinate of Fe$_{1}$ and $l$ is a Miller index.

We fit the ellipticity to the FLPA analysis of the
$[0,0,\tau]$ reflection in Fig.~\ref{fig:FeAs:Softpolanalysis} (top)
and the azimuthal measurement of the $[0,0,2-\tau]$ reflection shown
in Fig.~\ref{fig:FeAs:azimuthP09}, which
shows the predicted azimuthal dependences for perfectly circular
helical structure, and elliptical structures.  

A circular structure
gives a constant intensity as a function of azimuthal angle, as
expected.  
As the structure is made more elliptical, the azimuth
changes from a constant to a sinusoidally-changing intensity.  If the
magnetic structure has the long axis of the ellipse pointing down the
$b$-axis, then the azimuth goes through a minimum at 180$^\circ$.
However, we find that 
the azimuth of the $[0,0,2-\tau]$ reflection goes through a minimum at
around 157$^\circ$.  
In our magnetisation model, the azimuth
intensity can only go to zero at 0$^\circ$and 180$^\circ$ or
90$^\circ$ and 270$^\circ$ depending on whether the long axis of the
ellipse is along the $a$- or $b$-axis. To account for
this, the ellipse is allowed to rotate such that long- and short-axes
no longer point along the $a$- and $b$-axes.  

In order to fit the azimuth of the $[0,0,2-\tau]$, the long axis of
the ellipse has to be placed along the \textit{b}-axis and rotated by
$-22^\circ$, as shown in Fig.~\ref{fig:FeAs:azimuthP09}. 
As the magnetic structure is elliptical rather than circular, the form
of the FLPA of the $[0,0,\tau]$ should be
highly dependent on the azimuth at which the measurement is taken. 
In the FLPA of the
$[0,0,\tau]$ reflection, the rotation of the ellipse has a similar
effect to changing the azimuth position at which the calculation is performed. 
The effect of ellipticity on the FLPA measurement is
quite dramatic, and is shown in Fig.~\ref{fig:FeAs:Softpolanalysis}.
For example, for a circular magnetic structure,
the predicted $P_1$ remains negative for all incident angles. The FLPA
does not require a rotation of the
magnetic ellipse to fit the data, but the rotation can be accommodated
by correcting for a potential offset in the azimuth position.

Combining both the azimuth of the
$[0,0,2-\tau]$ reflection and the FLPA of
the $[0,0,\tau]$ reflection allows a fit requiring only three parameters, the
ellipticity, $S_{b}/S_{a}$, the rotation, $\zeta$ and the azimuth
offset of the polarisation analysis measurement $\psi_\textrm{flpa}$. 
The final fit results in an ellipticity of $\frac{S_b}{S_a}=
2.58 \pm 0.03$, which is far more substantial than the ellipticity of
1.15 proposed by the neutron experiment of Rodriguez \textit{et
  al.}\cite{rodriguez}. 
The azimuthal
measurement suggests a rotation of the ellipse of $-21.9\pm0.2^\circ$.
(As a consequence the azimuthal position of FLPA needs to be $11.0\pm0.2^\circ$.  This value of the
azimuthal offset from the \textit{b}-axis is within the experimental
uncertainty of mounting the sample.)
 The differences between our results and those derived from the
 neutron experiments \cite{rodriguez} probably result from the very
 different techniques used to estimate the magnitude of the
 ellipticity. The  study of Rodriguez {\it et al.}\ used the intensity profiles of
 16 relatively low intensity nuclear and magnetic reflections from
 polarised neutron diffraction measurements. Our study however has
 used the intensity variation of the azimuthal dependence of a
 resonantly enhanced x-ray magnetic satellite as well as the full
 linear polarised analysis of a separate magnetic satellite, which
 were combined and fitted with a model involving just three adjustable
 parameters.

\subsection{The effect of spin canting}
The $[0,0,1-\tau]$ reflection is predicted to have zero
intensity from the above structure factor.
In order to explain the origin of this
reflection, a spin helix along the $c$-axis with moments restricted to lie within the $a$-$b$ plane is not
sufficient. The phase difference
brought about by the two-orbit structure is the origin for the
predicted extinction of the $[0,0,1-\tau]$ reflection, and not the
direction of the magnetisation vector.  This means that changing the
magnetic structure to a cycloid or collinear spin density
wave will not change the extinction of $[0,0,1-\tau]$, whilst the
two-orbit structure remains.  It is also the case that adding a
canting in the $c$-direction will not contribute to a satellite
peak
as adding a $c$-axis component to the magnetic moment that
oscillates with a periodicity of $\tau$ will contribute
only to the satellite of allowed Bragg peaks.

 If we assume the existence of an
easy-axis for the magnetic moment tied to the crystal geometry, it is
reasonable to assume that this lies in the $a$-$c$
plane.  
The black line in 
Fig.~\ref{fig:feas:cryst}
shows an
example direction for the easy-axis, and the green and blue arrows
show the canting effect on the $a$ component of the moment
towards the easy-axis. For the moments on the Fe$_1$ and Fe$_2$ sites
that make up one orbit of the helix, the magnetic easy-axis on the
Fe$_2$ site will be a reflection by the $\sigma_x$ mirror plane of the
easy-axis on site Fe$_1$, resulting in a canting in the opposite
direction along the $c$-axis.   The same relationship holds for
the canting effects between the Fe$_3$, Fe$_4$ sites.  As Fe$_1$ and
Fe$_2$ are half a unit cell apart the oscillation in the
$c$-axis canting can be described by a co-sinusoid with a
periodicity of the unit cell, with a phase shift between the two
orbits, proportional to the difference in the $z$-component of the
positions of Fe$_1$ and Fe$_3$.    The resulting $c$-axis
component of the magnetic moment is dependent on both its position
along the $c$-direction in the unit cell and the position
around the magnetic helix, as only the $a$-axis component of
the moment experiences a canting effect.

 A $c$-axis component  of the $j$th magnetic moment is given by 
\begin{equation}
\boldsymbol{m}_{n,j}\left(\mathbf{r}_{n,j}\right)\cdot \mathbf{\hat{c}} =
\alpha_{a,c}\beta_{n}S_a\cos\left(\boldsymbol{\tau}\cdot\mathbf{r}_{n,j}-\psi_n\right),
\end{equation}
where
$S_a\cos\left(\boldsymbol{\tau}\cdot\mathbf{r}_{n,j}-\psi_n\right)$
is the magnitude of the $a$ component of the spin, and $\alpha_{a,c}$ is a
constant that is determined by the strength of the canting effect, and
$\beta_{n}$ takes the value $\pm1$ depending on the atomic site,
(i.e.\ Fe$_1$ and Fe$_3$ take the value +1 and Fe$_2$ and
Fe$_4$ the value $-1$). The specific direction of magnetic easy-axis is
included in the the value $\beta_{n}$. For example,  if the easy-axis is
perpendicular to the example shown in Fig.~\ref{fig:feas:cryst},
then $\alpha_{a,c}$ takes a negative value;  if the easy-axis
is entirely along the \textit{a}- or \textit{c}-axis then
$\alpha_{a,c}$ will be zero.  

Including the $c$-axis component in the structure factor allows
the simulation of the full polarisation analysis of the $[0,0,1-\tau]$
(solid lines in Fig.~\ref{fig:FeAs:Softpolanalysis}).
The predicted structure factor for the $[0,0,1-\tau]$ is only
dependent on the $c$-axis component of the magnetization
vector.  There are no parameters to fit in simulating the FLPA, as the only parameter $\alpha_{a,c}$, controls
the strength of the tilting, i.e.\ the magnitude of the
$c$-axis component with respect to the helical component, and
consequently has no impact on the polarisation dependence of the
$[0,0,1-\tau]$.  The simulation of the polarisation dependence of the
$[0,0,1-\tau]$ reflection gives good agreement  with a collinear
$c$-axis moment. This confirms the canted model, where only the
component of the moment in the $c$-direction experiences a
canting effect, and the Fe$_{1,3}$ and Fe$_{2,4}$ sites have opposite
canting effects.  However, the full polarisation analysis of the
$[0,0,1-\tau]$ reflection does not contain information about the
magnitude of the canting, while the $[0,0,\tau]$ reflection only contains
information about the $c$-axis component.
 We note that it is therefore not possible to extract the size of the canting from
polarisation analysis of reflections along the $[0,0,L]$ direction.
However, we note here that the results of our DFT measurements
described in Section~\ref{dft_section}, allow us to provide physical
motivation for the origin of the canting and therefore further
evidence that our model is appropriate.

\begin{figure}[!t]
\begin{center}
\includegraphics[width = \columnwidth]{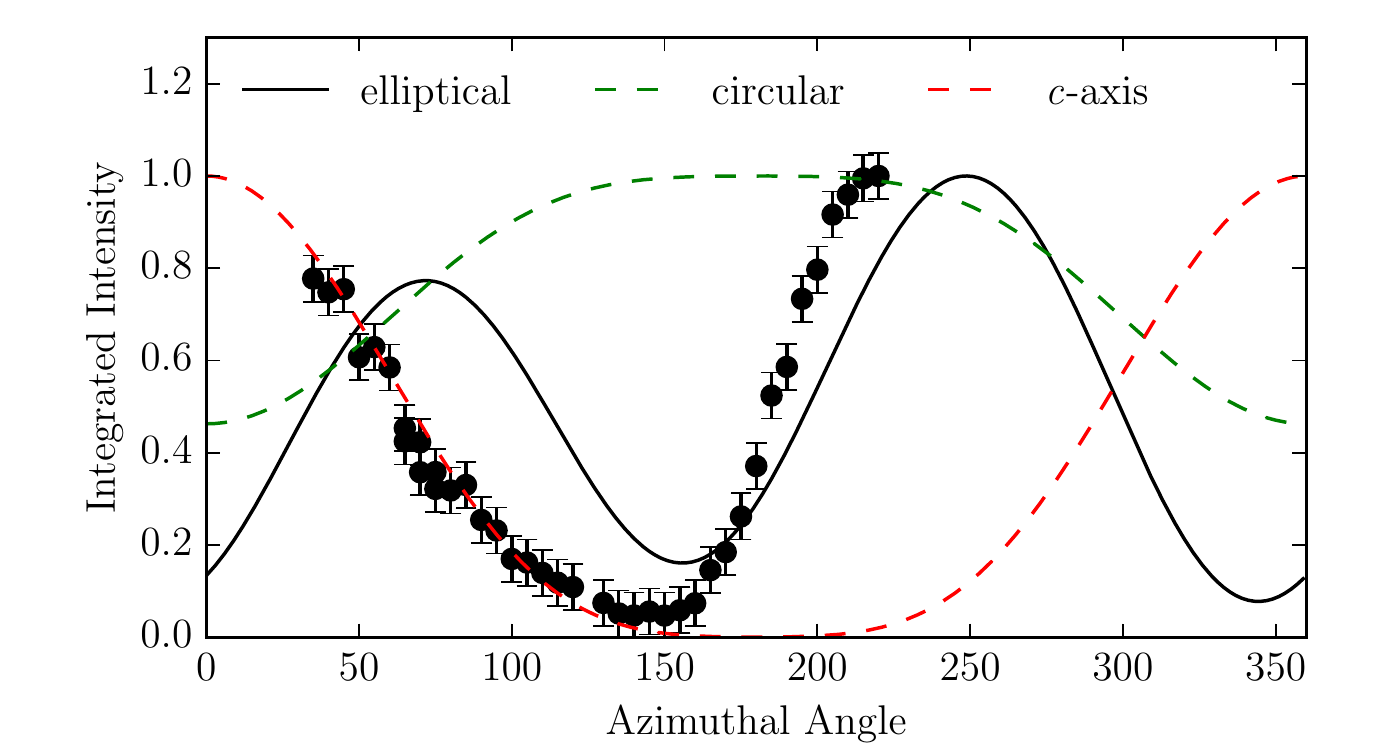}
\caption[Azimuthal measurement of the ${[1,0,3-\tau]}$ magnetic Bragg
peak]{Azimuthal measurement of the $[1,0,3-\tau]$ magnetic Bragg peak.
  The green dashed line shows the predicted azimuth for a circular
  helical magnetic structure.  The black solid lines shows the
  prediction for an elliptical rotated magnetic structure. The red
  dashed line shows the results predicted using a magnetic helix
  pointing along the \textit{c}-axis. These
  predictions were calculated using the structure factors.} 
\label{fig:FeAs:azimuthOAP09}
\end{center}
\end{figure}

Finally, the $[1,0,3-\tau]$ reflection was measured and an azimuthal dependence
of the scattering collected. The structure factor for this reflection
is dominated by the helical magnetic term, unless the phase difference
between the orbits falls within the range 1.8 and 2.0 radians, canting
strength dependent, in which case the canting component becomes the
more dominant.   Figure~\ref{fig:FeAs:azimuthOAP09} shows the results
of the azimuthal dependence of the $[1,0,3-\tau]$ satellite
reflection.  
Three different models are shown in
Fig.~\ref{fig:FeAs:azimuthOAP09}: the predicted azimuth for a
circular helical structure with spins restricted within the $a$-$b$
plane; 
a rotated elliptical structure; and the prediction using the
\textit{c}-component only.   The data does not agree quantitatively
with any of the three models,  while qualitatively it most resembles
the elliptical model.  The \textit{c}-component and circular models
are significantly different from the measured result.  This result is
sufficient to further rule out the non-elliptical case, but the
measurement is not sufficient to gain any information about the phase
difference between the two orbits, nor the magnitude of the canting.

\subsection{Determination of the chirality}

\begin{figure}[!t]
\begin{center}
\includegraphics[width = \columnwidth]{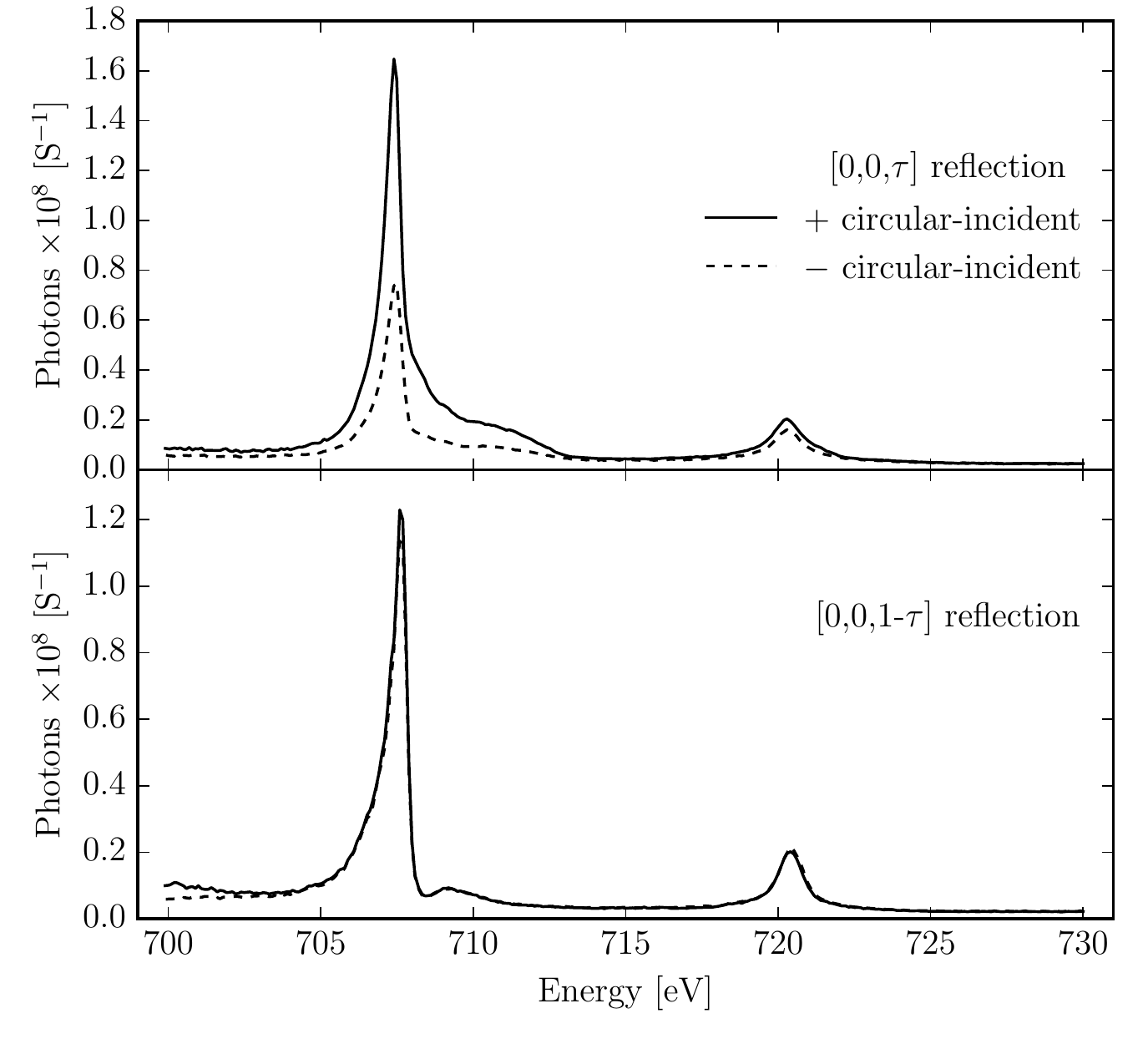}
\caption[Energy resonance of magnetic Bragg peaks at Fe
L${_\textrm{III}}$ (circular polarisation)]{Energy resonance of the
  magnetic satellite peaks, $[0,0,\tau]$ (top) and $[0,0,1-\tau]$
  (bottom), with circular-positive and negative-polarised incident
  light.}
\label{fig:FeAs:CIRenergySOFT}
\end{center}
\end{figure}

The chirality of the magnetic structure 
has no effect on the simulations for a full linear
polarisation analysis, but is important when circular incident light
is used\cite{mulders}.  For a chiral magnetic structure,
incident circular positive and incident circular negative light can be
used to establish the chirality\cite{mulders}. 

\begin{figure}[!t]
\begin{center}
\includegraphics[width = \columnwidth]{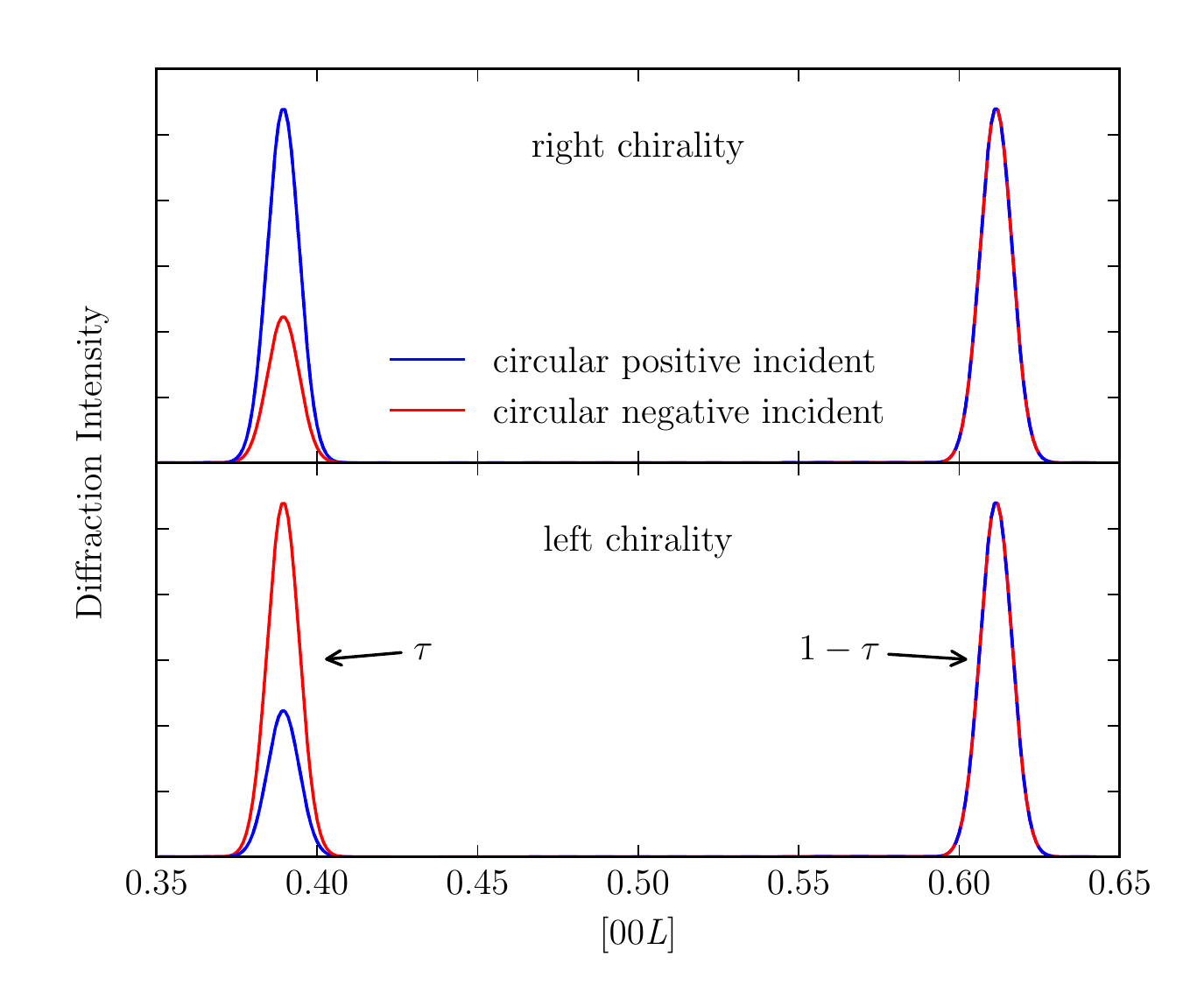}
\caption[Predicted intensities for circularly polarised light.]{Predicted intensities for circular incident polarisation for the $[0,0,\tau]$ and $[0,0,1-\tau]$ for both right chiral (top) and left chiral (bottom) helical magnetic structure.}
\label{fig:FeAs:predictedCIRC}
\end{center}
\end{figure}

Energy scans of the $[0,0,\tau]$ and $[0,0,1-\tau]$ reflections were
performed at the Fe $L_\textrm{II/III}$ edges 
(Fig.~\ref{fig:FeAs:CIRenergySOFT})
The predicted intensities of the
$[0,0,\tau]$ and $[0,0,1-\tau]$ peaks were made using the structure
factor calculations
for both chiralities
(Fig.~\ref{fig:FeAs:predictedCIRC}).  The intensity of the $[0,0,1-\tau]$
reflection does not change between circular positive and circular
negative incident light, as this peak is sensitive only to the
\textit{c}-axis component which does not have a chiral nature. The
$[0,0,\tau]$ peak does show a variation with incident circular light,
and predictions show that one circular channel is expected to be over
twice as intense as the other.  This is observed to be the case.  The
predictions also show that for a right-handed chiral helix  the
positive circular channel is expected to be the most intense, and for
a left-handed chiral helix the negative circular channel is expected
to be more intense.  The non chiral case, where the two orbits have
opposite chirality, is predicted to show equal intensity in the
circular positive and negative channels.  The energy scans show the
positive circular channel was the most intense channel, ruling out the
non-chiral case and strongly suggesting that the magnetic helix is
right-handed.

As a check, the linear polarisation analyser was used to examine the scattered
beam from the $[0,0,\tau]$ satellite reflection, with both circular
positive and circular negative incident light.  
Using the helical structure factor, with the parameters from the fit
of the linear polarisation and azimuth measurements, the analyser
scans were simulated for both chiral cases.  For the left and
right-handed chiral cases the positive and negative incident lights
are predicted to show opposite behaviour.  We find that our measurement resembles the
right-handed chiral structure, providing further confirmation that the magnetic helix
has right-handed chirality.

\begin{figure}[!t]
\begin{center}
\includegraphics[width = \columnwidth]{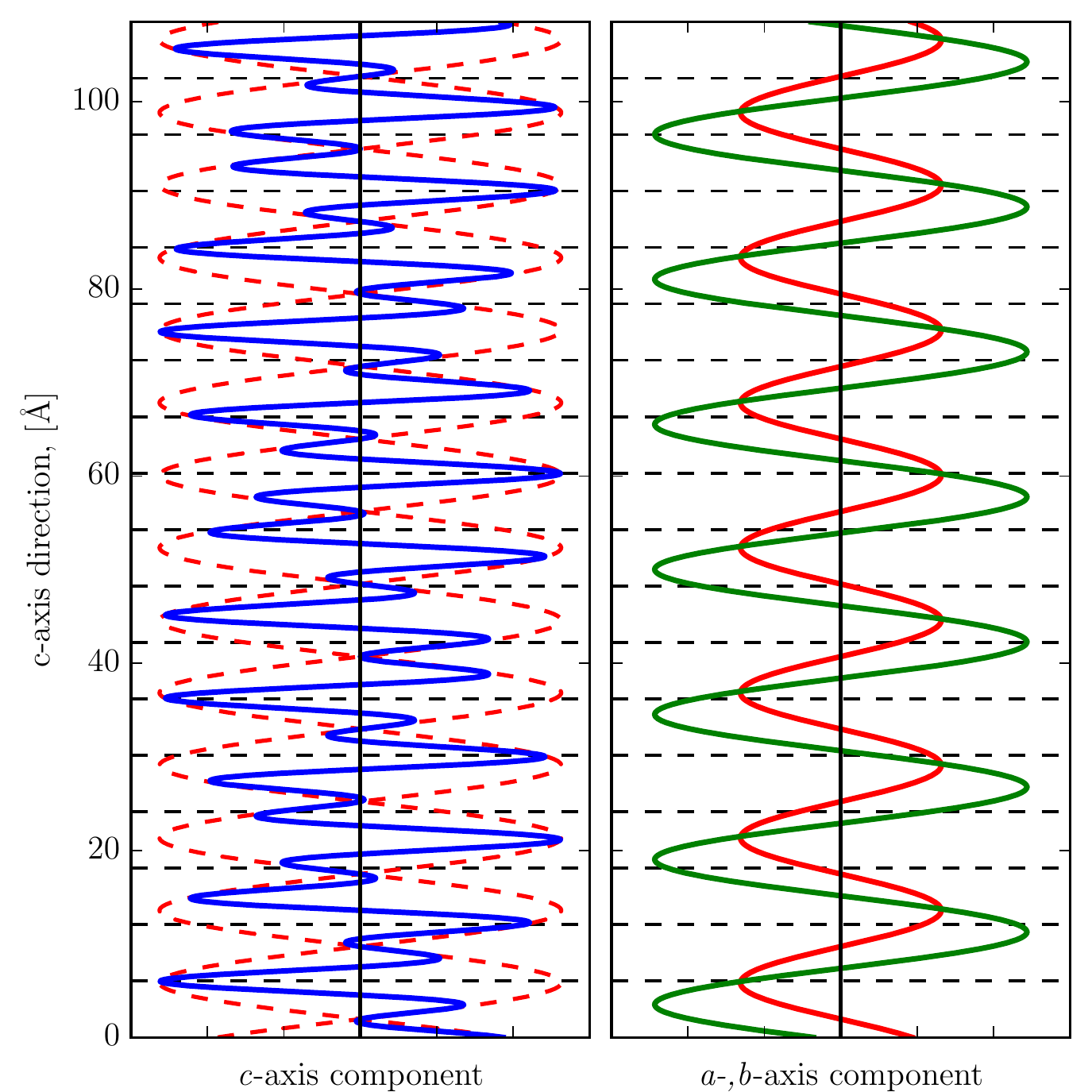}
\caption[Canted Helix]{\textit{a}-, \textit{b}-, and \textit{c}-axis
  components of the magnetic helix, shown in red, green and blue
  respectively.  The dashed red line shows the \textit{a}-axis component envelope around the \textit{c}-axis component.}
\label{fig:FeAs:magresult}
\end{center}
\end{figure}

The resonant x-ray scattering results have shown that a double circular
helical magnetic structure is insufficient, and that the helix maps
out an ellipse in the $a$-$b$ plane.  This ellipse has been shown to have an major axis
2.58 times longer than the minor axis.  The azimuthal measurement,
showed that major axis of the ellipses is rotated $-21^\circ$ away
from the $b$-axis.  The full polarisation analysis of the
unexpected $[0,0,1-\tau]$ peak, requires a canting of the
$a$-axis component of the moment into the $c$-direction
with a periodicity of the unit cell.  The absolute magnitude of the
canting cannot be found from the measurements taken, just its
presence.  The phase difference between the two magnetic orbits has
not been found.  Fig.~\ref{fig:FeAs:magresult} shows the
$a$-, $b$-, and $c$-axis components of the
magnetic moment for the canted spin helix, for one orbit.

This $a$-$c$ canting relation, in which the $a$-axis component of the
helix is canted with the
periodicity of the unit cell $c$ parameter, results in a total
magnetic structure with a periodicity longer than given by
$\tau$.  If we assign $\tau= 0.3\dot{8}$, such that the
commensurate position is $\frac{7}{18}$, then the effect of the canting is
to make the magnetic helix repeat every 18 unit cells along the
$c$-axis.  This can be seen in Fig.~\ref{fig:FeAs:magresult}
where the moment rotates around the helix seven times before returning
to its starting position. 
 It should be noted that we have assumed that the canting relation
is between the $a$- and $c$-moment directions, as canting in the
$b$-direction would break the reflection symmetry.
This unusual canting effect which only occurs along one direction of
the helix combined with the ellipticity explains the unusual magnetic
susceptibility measured by Segawa {\it et al.}\cite{segawa}. 
Whilst an elliptical helical structure goes some way to explain why
the susceptibility along the $b$-direction is lower than along
the $a$-direction, it does not explain the presence of the
field splitting in only one direction.

\section{DFT Calculations}\label{dft_section}

\begin{figure}
\begin{center}
\includegraphics[width = \columnwidth]{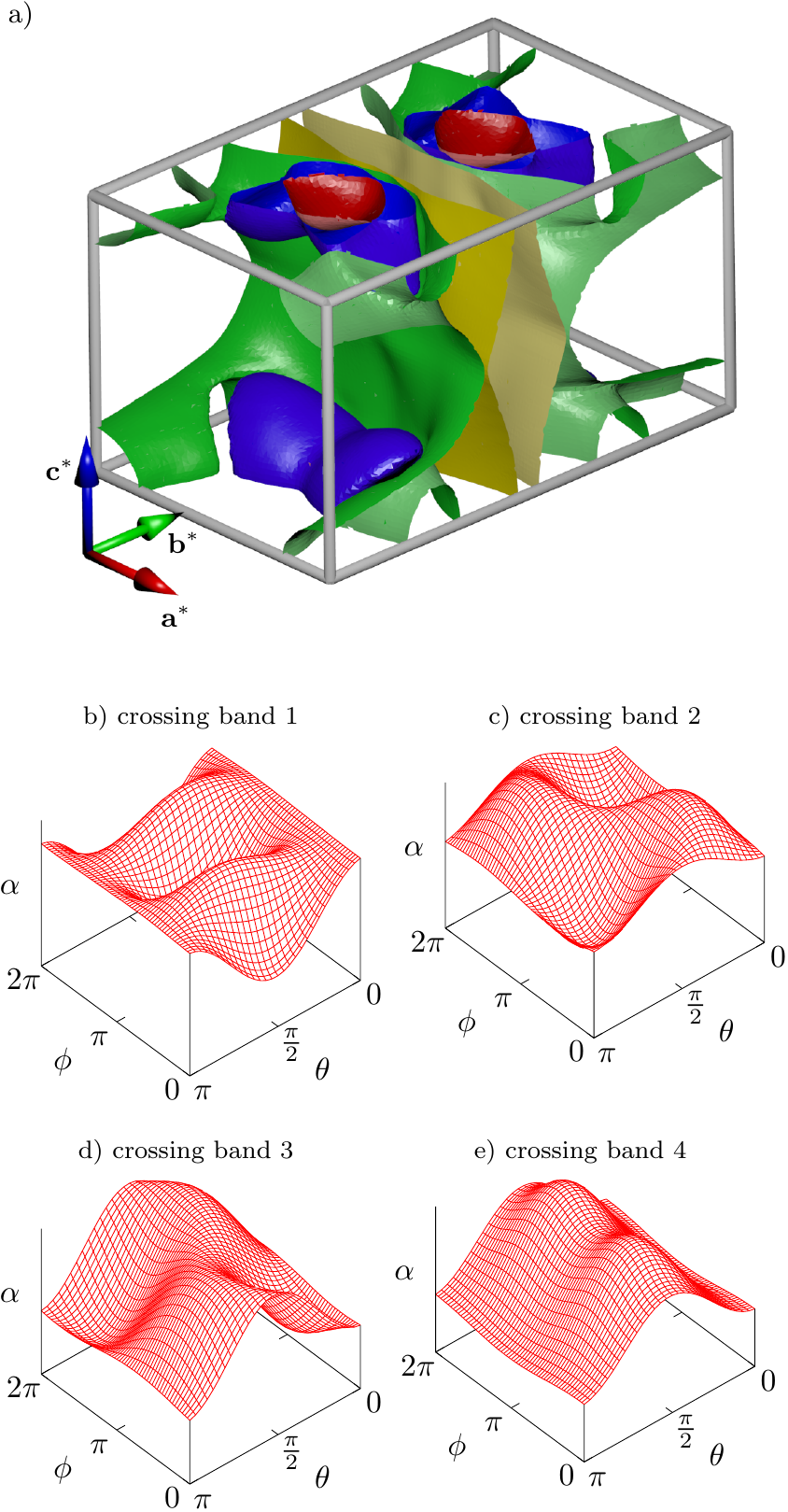}
\caption{(a) Fermi surface plot, (b)-(e) 2nd order
spin-orbit perturbation energy $\alpha$ for each Fermi
suface-crossing band, for different anglular directions $(\theta,\phi)$ of a Fe$_{1}$
probe spin. We use the  $Z$-$X$-$Z$ Euler angle conventions: $\theta$
is the angle away from $c$-axis, $\phi$ is the angle away from $b$-axis
after projecting into the $a$-$b$ plane. The axis coordinates
$(\theta, \phi)$ are given by
$a=((n+1/2)\pi, (p + 1/2)\pi)$, $b=((n + 1/2)\pi, p\pi)$,
$c=(n\pi, \phi)$ with $n$ and  $p$ integers. \label{fig:robert2}}
\end{center}
\end{figure}

Zero-spin DFT calculations on iron arsenide converged to a state with
four bands crossing the Fermi surface, as shown in
Fig.~\ref{fig:robert2}(a). Hole and electron
curvature is present at the Fermi surface, in agreement with
previous calculations and experiment \cite{mazin,saparov}. No
symmetry was enforced in the calculation and
both the LDA and the GGA calculations converged onto a $Pnma$
symmetry.

\begin{table}
\begin{center}
  \begin{tabular}{ | l || r | r | r | r | r | r |}
    \hline
    State & AFM1 & AFM2 & AFM3 & FMSF & FM & zero-spin \\ \hline
    Rel. energy meV/Fe & 0 & 19 &25& 25 & 50 & 100 \\ \hline
    Spin mag. $\mu_\mathrm{B}$/Fe & 1.32 & 1.06 &1.00 & 0.98* & 0.6 & 0 \\ \hline
  \end{tabular}
  \caption{Relative energies and spin magnitudes for different states calculated with the GGA, which show an almost linear relation between the two. *FMSF state has varying spin magnitudes, the mean value is stated here.}
  \label{tab:PBEres}
\end{center}
\end{table}

Collinear spin-polarized calculations were performed
for all spin parallel-antiparallel pairings in the unit cell. There
are three antiferromagnetic states, a ferromagnetic state (FM), and a
ferrimagnetic state where one spin is flipped from the FM state (FMSF). The
antiferromagnetic states are identified by the iron atoms which have
parallel spins: $\mathrm{Fe}_1 \parallel \mathrm{Fe}_3 $ (AFM1),
$\mathrm{Fe}_1 \parallel \mathrm{Fe}_4 $ (AFM2), and
$\mathrm{Fe}_1 \parallel \mathrm{Fe}_2 $ (AFM3). The relative energies
and ordered spin moment for these states and the zero-spin state
using the GGA are shown in Table~\ref{tab:PBEres}. These agree
with previous calculations\cite{mazin}. The LDA results
follow the same trends as the GGA results, but with lower ordered
moments. 

The magnitude of the  ordered moment is found to increase with the
number of antiferromagnetically aligned pairs of Fe moments. 
We also  find that the the energy of the
states varies linearly with ordered moment.
The energy relative to the AFM3 state is best fitted by $E=
\gamma \sum_i |s_i|+\beta$, where the sum is over spins $i$,  $\gamma=-74.8$~meV/$\mu_{\mathrm{B}}$
and $\beta=394$~meV. 
The observed linear energy dependence on spin magnitude contrasts sharply with
the Heisenberg and Ising models which have a quadratic energy
dependence, and is instead reminiscent of a Stoner instability as
found in ferromagnetic metals\cite{stoner_FM}. 
Examination of the
electron density in the system also shows that it changes with the
transition to the ordered spin-state.
To explore the origin of this linear dependence we calculated
differences in total electron density between states. Total density
differences between the zero-spin state  and AFM1 state 
are shown in Fig.\ref{fig:robert1}(c) and (d). These show that when the
system is relaxed into the AFM1 state, the electron density descreases between iron
nearest-neighbours (Fe$_1$-Fe$_2$, Fe$_3$-Fe$_4$) and increases
along Fe-As bonds (Fe$_n$-As$_n$) in a pattern that forms strings
along the $a$-direction. The arsenic atoms were found to have a 
complicated local arrangement of spin density
with no overall magnetic moment. 

From our calculations we are able to provide an insight into
the $a$-$c$ canting relation
of the magnetic order arises. 
 The
helical magnetic ordering wavevector is found to  change with temperature \cite{rodriguez}, and
this could imply that the system is sensitive 
to small perturbations.
In fact, there are several frustrated non-equivalent Fe-As bonds in
the material which are
likely to be responsible for the
sensitivity of the ground state.
There are three inequivalent Fe-Fe bonds, although one bond
(Fe$_1$-Fe$_3$) is seperated by a much larger distance than the
others.
We find that the spin moment magnitude on individual iron sites is correlated with the
number of antiferromagnetically aligned nearest neighbour Fe atoms.
In our
calculations, an iron atom with no antiferromagnetic short bonds has
a moment of $0.6\mu_{\mathrm{B}}$;
those with two antiferromagnetic short bonds have  a moment of 
$\approx 1\mu_{\mathrm{B}}$, 
and those with all four antiferromagnetic short bonds  have a
magnitude of $1.32\mu_{\mathrm{B}}$. In
the lowest energy state (AFM3), iron atoms linked by a long bond
(Fe$_1$-Fe$_3$) must be ferromagnetically aligned.  
The trend in Table I shows that the structures with more
antiferromagnetic bonds have a larger spin magnitude and lower total
energy. 
An explanation for the occurence of the helical magnetic
 state rather than AFM3 
is that it allows the energetically unfavourable ferromagnetic pairing
along the long bond in the AFM3 structure to reduce energy by canting.

Due to the 
sensitivity of the ground state, 
we expect weak mechanisms such as spin-orbit coupling to
be decisive in realizing the ground state magnetic structure, 
and here we show that spin-orbit coupling explains the
presence of the ordered component in the $c$-direction.
In order to see the effect of the spin orbit interaction on the
ordered magnetic structure, 
we calculate the energetic perturbation of the spin-orbit interaction
between the iron electron spins and the projected atomic orbitals.
This is
used to estimate the preferred direction of Fe spin alignment. 

The perturbation to the ground state energy can be calculated using the
minimization of the energy density functional given by
\begin{equation} 
H=H_{0}[\rho]+\varepsilon H_{\mathrm{spin\text{-}orbit}} [\rho],
\end{equation}
where $H_{0}[\rho]$ is our unperturbed density functional for the
energy and $\rho$ is the electron density. In the Kohn-Sham representation we can use the usual
electronic formulation for the spin-orbit interaction, including it as
a correction term to the electrostatic field around a given atom. This
is given by 
\begin{equation}
H_\mathrm{spin\text{-}orbit}=- \frac{\mu_\mathrm{B} }{e \hbar   c^2 }  \sum_\Psi \langle \Psi| ~
\mathbf{s} \cdot
\mathbf{L}\frac{\mathrm d V}{\mathrm d r_{\mathrm{atom}} }
\frac{1}{r_\mathrm{atom} m}| \Psi\rangle,
\end{equation}
where the sum is over the Kohn-Sham orbitals $\Psi$, $V$ is the electrostatic potential around the atomic nucleus, $r_\mathrm{atom}$ is the
distance to the centre of the atom, and $m$ is the band mass of the KS
orbital.
We approximate the electric field as radial,
and this is valid as long as the core orbitals that shield the nuclear
charge are not affected by the bonding. 
This radial symmetry leads to
the condition $\left[\mathbf{L},\frac{\mathrm d V}{\mathrm d r_{\mathrm{atom}} }\right]=0$, and further
for FeAs we only consider the Fe $3d$ orbitals which are involved in the
magnetic structure.
This permits a significant simplification of
the calculation. 

For zero-spin calculations where
$\rho_{\uparrow}=\rho_\downarrow$,  the energetic perturbation
$\alpha$ is
second-order in the perturbation parameter\cite{sakurai} $\varepsilon$, and
is given by
\begin{equation} \label{equation:fincalc}
\alpha\propto -\varepsilon^2 \int \frac{
   \left|\langle \Psi |L_{z'}/m | \Psi\rangle \right|^2
 }{ |\nabla U|} \mathrm d \mathbf \Psi, 
\end{equation}
where $z'$ is the direction of the spin moment on an iron atom and
$L_{z'}$ is the component of angular momentum in this direction. 
The integral is over the orbitals $\Psi$ evaluated at the Fermi surface, 
and is dependent on the gradient of the energy $U$ of the Kohn-Sham orbitals
at the Fermi surface $ |\nabla U (\Psi)|$. We find that 
the diamagnetic contribution of the orbitals is small compared to
$\alpha$ and has therefore been neglected. 
A perturbation to the lowest nonzero order does
not change 
the electron density from that of the ground state, and so 
we are free to choose the  value of $z'$ and calculate a physically meaningful energetic
perturbation for this chosen spin orientation \cite{janak}. 
We are therefore able to create a full map of the energetic
perturbation 
for different
spin alignments. This allows us to 
assess the spin anisotropy of a specific Fe atom.

The principle parts of a plane-wave pseudopotential calculation are the
projections of the Kohn-Sham orbitals onto the atomic basis
set. 
The atomic orbital
projections do not necessarily obey crystal symmetries, 
To generate the full set of projections the relevant local symmetry
operators are calculated from the Wigner-$d$ matrices
\cite{vn2336439} and, if required, the application of a
reflection. From this complete orbital projection, projection
amplitudes at the Fermi surface are calculated using a $B$-spline
interpolation\cite{Einspline}. This
projection of the Fermi surface is then used as the basis for the
energetic perturbation computation. 

The final result is calculated by applying Eq.~\ref{equation:fincalc}
to each Fermi surface point, and 
$L_{ z'}$ is calculated using a Mulliken orbital
projection \cite{MullikenPop}, and the use of Wigner-$d$ matrices to
include a rotation from the  $z$-direction to $z'$\cite{vn2336439}. This is
performed successively for each value of $z'$ to generate a full map
(in energy) of
the perturbation, which is chosen to be a polar map with regular
intervals in both $\theta$ and $\phi$ coordinates.
We take the unperturbed state (corresponding to $H_{0}[\rho]$) to be the zero spin
configuration. 
This state was
chosen for the calculation 
as it has the largest Fermi
surface. 
Results were obtained for an Fe$_{1}$ test spin  aligned along 
different directions, and these energies are shown, for each band
in Fig.~\ref{fig:robert1}. 
The calculations show clearly the effects of anisotropy, which causes a
large difference in energy for different iron spin orientations. 
In general, we note that the extrema in energy do not lie along a crystal axis.

Crossing bands 3 and 4 (Fig.~\ref{fig:robert1}) makes a significantly
larger contribution to $\alpha$
than the other bands, as they have the largest Fermi
surfaces and the highest density of $d$-orbitals.
 On  band 3 the
highest energy perturbation occurs when the Fe spin points in the
$a$-$c$ plane, at an angle of 23$^{\circ}$ from the $c$-direction
(towards $a$).  We find that
spin alignment along the $b$-axis is energetically unfavourable. The
results for band 3 compare favourably with the measured
susceptibilities in the high temperature paramagnetic spin state, in
which $\chi_a \approx \chi_c>\chi_b$ \cite{segawa}. 

Finally, the
anisotropy of the Fe $d$-orbitals is calculated here from the
second order correction energy $\alpha(\theta,\phi)$ as
$(1-\alpha_{\mathrm{min}}/\alpha_{\mathrm{max}})$. This quantity, on
band 3, is high in the $a$-$c$ 
plane, at 81\%, meaning that the local environment
strongly affects the Fe $d$-orbitals in this band. The other bands 
have  anisotropies 77\% (band 1), 51\% (band 2)
 and 97\% (band 4). 
For crossing band 2 the lowest energy spin direction lies exactly midway between
$a$ and $-c$; and in crossing band 4 it lies along $c$. 
The moment on the iron cannot satisfy all of these conditions
simultaneously, but by far the largest proportion of $d$-orbitals lie on
crossing band 3. 
However,  it is notable that 
the optimal direction of spin alignment lies off-axis in the
$a$-$c$ plane, and that spin-orbit effects will
couple ordering in the $a$ and $c$ directions, with
the relative orientation dependent on the iron site.

\section{Conclusions}

In conclusion, 
we have used polarised resonant x-ray scattering
measurements to investigate the incommensurate nature of the magnetic
helix  along the $c$-axis
in FeAs as well as its ellipticity. We have found evidence of a much
greater ellipticity to that inferred previously,  
as well as a $a$-$c$ canting relation in which moments are canted out
of the $a$-$b$ plane
 In addition by
use of circular polarised x-rays we have demonstrated the existence of
a right-handed chiral structure. We have combined our experimental
measurements with DFT calculations which have quantified the relative
energies of different antiferromagnetic states and showed   that the origin of
the spin canting effect we have measured may be accounted for by considering the
spin-orbit coupling. 
Finally, we note that the observation of
both the fundamental magnetic satellite and also higher-order
harmonics in FeAs suggest a different behaviour to modulated
antiferromagnetism in elemental metals such as chromium and the
rare-earths.

\section{Acknowledgements}
We are grateful to EPSRC and the John Templeton Foundation for
financial support. 
We acknowledge Diamond Light Source for time on beamline I10 under
proposal SI6422. Parts of this research were carried out using PO9 at the light
source PETRA III at DESY, a member of the Helmholtz Association (HGF)
(proposal I-20110275 EC).
Work at King Fahd University of Petroleum and Minerals was supported
by National Science Technology and Innovation Plan (NSTIP)  under
project No.\ 11-ADV1631-04.
We acknowledge the UK national supercomputing facility (Archer),
Durham HPC (Hamilton), the facilities of N8 HPC and the UKCP
Consortium (EPSRC Grant No. EP/F037481/1).
Data presented in this paper will be made available via
doi.org/10.15128/r36h440s444.

\end{document}